\documentclass[conference]{IEEEtran}
\IEEEoverridecommandlockouts
\usepackage{cite}
\usepackage{amsmath,amssymb,amsfonts}
\usepackage{algorithmic}
\usepackage{graphicx}
\usepackage{textcomp}
\usepackage{xcolor}
\usepackage{verbatim}
\usepackage{hyperref}
\usepackage{listings}
\usepackage{tabularx}
\usepackage{arydshln}
\usepackage[frozencache]{minted}
\usepackage{adjustbox}
\usepackage{booktabs}
\def\BibTeX{{\rm B\kern-.05em{\sc i\kern-.025em b}\kern-.08em
    T\kern-.1667em\lower.7ex\hbox{E}\kern-.125emX}}

\begin{document}

\title{RDMA vs.~RPC for Implementing Distributed Data Structures
%\thanks{Identify applicable funding agency here. If none, delete this.}
}

\author{
  \IEEEauthorblockN{
    Benjamin Brock\IEEEauthorrefmark{1}\IEEEauthorrefmark{2},
    Yuxin Chen\IEEEauthorrefmark{3}\IEEEauthorrefmark{2},
    Jiakun Yan\IEEEauthorrefmark{4}\IEEEauthorrefmark{2},
    John D. Owens\IEEEauthorrefmark{3}\IEEEauthorrefmark{2},
    Ayd\i{}n Bulu\c{c}\IEEEauthorrefmark{1}\IEEEauthorrefmark{2},
    and Katherine Yelick\IEEEauthorrefmark{1}\IEEEauthorrefmark{2}
  }
  \IEEEauthorblockA{
    \IEEEauthorrefmark{1}University of California, Berkeley\\
    \IEEEauthorrefmark{2}Lawrence Berkeley National Laboratory\\
    \IEEEauthorrefmark{3}University of California, Davis\\
    \IEEEauthorrefmark{4}Shanghai Jiao Tong University\\
  }
}

%\author{

%\IEEEauthorblockN{Anonymous Authors}
%\IEEEauthorblockA{\textit{dept. name of organization (of Aff.)} \\
%\textit{name of organization (of Aff.)}\\
%City, Country \\
%email address}

%\IEEEauthorblockN{Benjamin Brock}
%\IEEEauthorblockA{\textit{dept. name of organization (of Aff.)} \\
%\textit{name of organization (of Aff.)}\\
%City, Country \\
%email address}
%\and
%\IEEEauthorblockN{Yuxin Chen}
%\IEEEauthorblockA{\textit{dept. name of organization (of Aff.)} \\
%\textit{name of organization (of Aff.)}\\
%City, Country \\
%email address}
%\and
%\IEEEauthorblockN{Jiakun Yan}
%\IEEEauthorblockA{\textit{dept. name of organization (of Aff.)} \\
%\textit{name of organization (of Aff.)}\\
%City, Country \\
%email address}
%\and
%\IEEEauthorblockN{John D. Owens}
%\IEEEauthorblockA{\textit{dept. name of organization (of Aff.)} \\
%\textit{name of organization (of Aff.)}\\
%City, Country \\
%email address}
%\and
%\IEEEauthorblockN{Ayd\i{}n Bulu\c{c}}
%\IEEEauthorblockA{\textit{dept. name of organization (of Aff.)} \\
%\textit{name of organization (of Aff.)}\\
%City, Country \\
%email address}
%\and
%\IEEEauthorblockN{Katherine Yelick}
%\IEEEauthorblockA{\textit{dept. name of organization (of Aff.)} \\
%\textit{name of organization (of Aff.)}\\
%City, Country \\
%email address}
%}

\maketitle

\begin{abstract}
Distributed data structures are key to implementing scalable
applications for scientific simulations and data analysis.  In this paper we
look at two implementation styles for distributed data structures:
remote direct memory access (RDMA) and remote
procedure call (RPC).  We focus on operations that require
individual accesses to remote portions of a distributed data structure, e.g.,
accessing a hash table bucket or distributed queue, rather than global
operations in which all processors collectively exchange information.
We look at the trade-offs between the two styles through microbenchmarks and
a performance model that approximates the cost of each.  The RDMA
operations have direct hardware support in the network and therefore
lower latency and overhead, while the RPC operations are more
expressive but higher cost and can suffer from lack of attentiveness
from the remote side.   We also run experiments to compare the real-world performance
of RDMA- and RPC-based data structure operations with the
predicted performance to evaluate the accuracy of our model, and show
that while the model does not always precisely predict running time, it
allows us to choose the best implementation in the examples shown.
We believe this analysis will assist
developers in designing data structures that will perform well on
current network architectures, as well as network architects in
providing better support for this class of distributed data structures.
\end{abstract}


% A note on terminology:
% I use the word `RPC` here, instead of `AM`, because
% I think it will be understandable by a broader
% audience.  For the purposes of this paper, we can
% consider AMs to be a restricted type of RPC.

\section{Introduction}
Many complex programs need to perform operations on abstract data structures, such as hash tables, queues, and arrays.   While many mature, high quality libraries exist that provide implementations of abstract data structures for serial and multi-threaded programs, the development of techniques for high-level data structures for distributed programs is still an active area of research
\cite{Aguilera:2019:DFM:3317550.3321433, Brock:2019:BCD:3337821.3337912, Fuerlinger:2016:DASH}.
Of particular interest are distributed data structures for \textit{irregular applications}, where data access patterns and volumes are not known in advance.  These applications commonly use data structures which may be complex to implement using traditional message passing methods in a distributed memory setting, including graphs, trees, hash tables, and distributed queues.
% TODO: more cites
Some recently developed distributed data structure libraries are founded on \textit{remote direct memory access} (RDMA), meaning that all essential data structure operations will be executed using one-sided remote put, get, and atomic operations~\cite{Brock:2019:BCD:3337821.3337912, Fuerlinger:2016:DASH}.  These data structure operations have the potential to be very efficient and to offer low latency, since they operate directly on remote data structure elements and can be executed directly by the network interface card (NIC) on most modern supercomputer and datacenter systems.  Other high-level programming environments encourage users to use \textit{remote procedure call} (RPC) software primitives to build distributed data structures~\cite{bachan2019upc++}. While RPCs require the attention of a remote CPU, which leads to higher latency, they have the potential to be much more expressive than the RDMA operations available on today's interconnects, potentially leading to fewer round trips.

In this paper, we evaluate the efficacy of RDMA- and RPC-based manipulation of distributed data structures with a set of systematic benchmarks.  We perform two sets of experimemts.  First, we perform microbenchmarks to gather the costs of the \textit{component operations} that make up both RDMA- and RPC-based data structure implementations.  This includes the cost of various RDMA operations along with the cost of an RPC.  Second, we measure the actual costs of various \textit{data structure operations}, such as queue or hash table insertions, at various levels of concurrency requirements using both RDMA and RPC-based implementations.  We then compare the observed results with an analytical cost model and determine where it is more desirable to use RDMA and where it is more desirable to use RPC.

We break down the cost of RDMA-based data structure operations in terms of an \textit{analytical cost model}, which we use to predict the cost of RDMA-based data structure operations based on the real measured cost of the component operations.  This type of analysis helps us to determine \textit{why} RDMA-based operations are expensive, when they are expensive, and to focus in on (1) where data structures could be improved to run on current-generation network hardware (e.g. by avoiding expensive operations), and (2) which operations hardware designers might focus on optimizing in order to better support distributed data structures.  Our paper has three main contributions.

\begin{itemize}
    \item We present a set of microbenchmarks that determine the cost of various component operations for RDMA-based distributed data structures on a modern supercomputer network.
    \item We present an analytical cost model which can be used to estimate the cost of RDMA-based distributed data structure operations, based on the component costs.
    \item We provide a comparison of RDMA- and RPC-based distributed data structure performance for queue and hash table data structures at variable levels of concurrency requirements.
    % \item We explore the discrepancy between the real measured performance and the predicted performance of the analytical cost model and conclude several reasons, which future designers could pay attention to for better performance.
    % TODO: Possibly more???
\end{itemize}

\section{Background}
\subsection{Remote Direct Memory Access}
Remote direct memory access (RDMA) provides an interface to manipulate remote data in a \textit{one-sided} manner, meaning that an origin process can perform operations on the remote memory of a target process without any explicit coordination with the target.
This is commonly executed by having the target's network interface card (NIC) directly communicate with its on-node memory, resulting in very low round-trip latency on the order of a microsecond.  Low-latency RDMA primitives are now available on a number of supercomputer interconnects, including Cray Aries and Infiniband.  RDMA is also increasingly available on datacenter commodity hardware through RDMA over converged Ethernet (RoCE).
% TODO: check RoCE spelling
% TODO: and others?

For the purposes of this paper, we consider a common set of RDMA operations available in most modern supercomputer and datacenter systems.  This set includes remote put and remote get, which can be of variable size, along with the fixed-size 32 and 64-bit atomic memory operations (AMOs) compare-and-swap and fetch-and-op.  Some have proposed an expanded set of RDMA operations to support various types of remote and distributed data structures, such as the Infiniband extended atomics API~\cite{mellanox_extended}.  In addition, there are recently proposed API extensions to RDMA which would allow for more expressive RDMA operations~\cite{Aguilera:2019:DFM:3317550.3321433}.  These APIs are outside the scope of this paper.
% TODO: add Gen-Z, etc. reference?

% TODO: limitations of RDMA

\paragraph*{RDMA-Based Data Structures}
\label{sec:pgas_atomicity}
Distributed data structures can be directly built on top of one-sided RDMA operations, so that all major data structure operations will be executed with RDMA.  Examples of such partitioned global address space (PGAS) distributed data structure libraries include BCL, DASH, and
Multipol~\cite{Brock:2019:BCD:3337821.3337912,Fuerlinger:2016:DASH,chakrabarti1995multipol}.
Similar to shared memory concurrent data structures, these libraries are built to use a shared global memory space, with synchronization using atomics when necessary, to operate upon shared data.  However, unlike shared memory data structures, the component costs and synchronization models of distributed programming frameworks can be quite different, so care must be taken to design data structures accordingly.  As shown in Figure~\ref{fig:pgas_ds}, libraries can use RDMA operations, which will be directly executed by the target process' NIC, to operate on remote data.  There are two remote memory operations in this code example, \texttt{CAS}, which is a remote compare-and-swap operation, and \texttt{RPUT}, which is a remote put operation.  In the best case, our inserting process will perform a remote compare-and-swap, succeed in reserving the first hash table slot, and then perform a remote put operation.  This would have a cost of $A_{\text{CAS}} + W$, that is the cost of a compare-and-swap operation and a write.  However, in the case of hash table collision, the algorithm will move on to the next available slot, and multiple round trips may be required to perform the insert operation.  The particular hash table shown here is a hash table with open addressing and linear probing.  Observant readers of Figure~\ref{fig:pgas_ds}  will also notice that the listed code is not fully atomic.  While the code is atomic with respect to concurrent insert operations, there is no guarantee that the remote put operation will finish before a remote find operation reads the half-written value.  If we wish for our insert operation to be atomic with respect to concurrent find operations, we will require a second fetch-and-op operation to mark the slot as ready for reading after the remote put operation has finalized.  This would increase the best case cost of the remote insert operation to $A_{\text{CAS}} + W + A_{\text{FAO}}$.  So, depending on an application's atomicity requirements, data structure operations over RDMA may have different best-case costs.  Also, depending on a particular execution of the application, the observed cost of a method may vary due to the number of round trips caused by contention.

% TODO: change this to minted.
\begin{figure}
\begin{minted}{Cpp}
void insert(key, val) {
  slot = hash(key);
  while (!success) {
    rval = CAS(flags + slot,
               free_flag, taken_flag);
    success = (rval == free_flag);
    if (!success) {
      slot++;
    }
  }
  if (success) {
    RPUT(data + slot, {key, val});
  }
}
\end{minted}
  % \lstinputlisting[language=C++]{pgas_ds.cpp}
  \caption{Modifying a hash table using one-sided RDMA operations.}
  \label{fig:pgas_ds}
\end{figure}

% Remote direct memory access (RDMA) allows one node to directly access the memory of a remote node without involving CPU on the remote side. This enables zero-copy transfers, reducing latency and CPU overhead. RDMA requires hardware support, and in our experiments, we use InfiniBand, which is a switched fabric network widely used in high performance computing systems. InfiniBand achieves low latency by implementing several layers of the network stack (transport layer through physical layer) in hardware, and by providing RDMA and kernel-bypass. Existing systems can transfer 1 KB in 1us using RDMA over InfiniBand FDR 4x. The semantics of RDMA memory operation are RDMA read (get), RDMA write (put) and RDMA atomic operations (compare and swap, add, etc). These operations are one-sided: the responder’s CPU is unaware of the operation. This lack of CPU overhead at the responder makes one-sided verbs attractive.

\subsection{Remote Procedure Calls}
Remote procedure calls (RPCs), in contrast to RDMA  operations, allow an origin process to remotely trigger the execution of a procedure on a target process.  RPCs have the advantage of being more expressive than RDMA.  While control flow in individual RDMA operations is limited to single-instruction atomics like compare-and-swap and fetch-and-op, RPCs can include complex control flow and arbitrary computation.  This allows more complex data structure operations, such as inserting into a hash table, pushing a value onto a queue, or even modifying a dynamically sized data structure, to be performed with a single communication event.  However, this added expressivity comes at a greater latency cost, since an RPC operation must wait for the target process to enter a progress function or interrupt the processor and make a function call to execute the procedure. It also changes load balancing across processors, moving away from the clearly defined SPMD model of execution in ways that can shift computational workload, intentionally or not.

% TODO: double-check if this definition is *strictly* what GASNet-EX's restrictions are.
% (I think this is actually slightly *more* restrictive.)
In this paper, we consider a restricted type of RPC called an active message (AM) \cite{vonEicken:1992:AMM:139669.140382}.  For the purposes of this paper, AMs have the following restrictions: (1) active message handlers may not send additional active messages, except for a single response to the origin process and (2) active message handlers may not perform network communication. These restrictions allow for a high-performance, low-latency implementation of active messages with bounded buffer space~\cite{bonachea2017gasnet,gasnet-lcpc18}.

\paragraph*{RPC Data Structures}
An implementation of a distributed data structure operation with RPCs requires two parts: (1) a handler function, which is the procedure that will be executed on the target process, and (2) a wrapper function, which is the function directly called by the user on the origin process and the code that will issue the RPC request.  RPC data structure implementations can be quite simple, as shown in Figure~\ref{fig:rpc_ds}.  The wrapper function \texttt{insert} uses a hash function to map data to the appropriate nodes, then issues an RPC with the handler function \texttt{insert\_handler}.  The handler function in this case simply inserts the key and value into a local hash table.  In contrast to the hash table implementation based on RDMA communication, this implementation will typically only require a single round trip over the network, since the origin node can push the RPC request onto the target node's RPC queue in a single network operation, then the target node can execute the necessary control flow to unpack and store the data.
% Depending on the payload size and RPC queue status, multiple round trips might be required to negotiate the RPC queue insertion.
% Removed at Dan's request, since this is not true of all GASNet conduits.
However, crucially, the number of network operations is unrelated to the control flow logic inside the data structure operation, which takes place on the target side inside the RPC function.  Depending on the specific manner in which the handler function will be called (either serially or simultaneously with other threads), the handler function may require local atomic operations or other mechanisms for synchronization.  However, these local mechanisms are significantly cheaper than remote memory operations.

\begin{figure}
  % \lstinputlisting[language=C++]{rpc_ds.cpp}
\begin{minted}{Cpp}
void insert_handler(key, val) {
  local_hash.insert({key, val});
}

void insert(key, val) {
  node = hash(key) % nprocs();
  rpc(node, insert_handler, key, val);
}
\end{minted}
  \caption{Modifying a hash table using an RPC.}
  \label{fig:rpc_ds}
\end{figure}

One important detail not directly illustrated by the above code listing is that the execution of the handler function is dependent on the attentiveness of the target process, which must enter a progress function in order for its RPC queue to be serviced.  While the liveness of RDMA operations is guaranteed by the network interface card, which will be constantly servicing instructions regardless of CPU state, RPC-based systems must either dedicate specific resources, such as a progress thread, to ensure attentiveness, or else pay the possible latency cost associated with waiting until the target process finishes its computation and enters a call to the RPC progress function.

\subsection{The Berkeley Container Library}
In this paper, we compare the performance of RDMA-based implementations of distributed data structures to RPC-based implementations.  For the RDMA-based implementations, we will benchmark data structures provided in the Berkeley Container Library (BCL).  BCL is a cross-platform library of distributed data structures that supports running on top of MPI, OpenSHMEM, and GASNet-EX.  BCL is a header-only library and is designed to offer high-level interfaces without any runtime cost for abstraction.  BCL data structures are built using remote put, remote get, and remote atomic operations such as atomic compare-and-swap and fetch-and-op.

\paragraph*{Performance Model}
Data structure operations in BCL can be characterized in terms of an analytical cost model, which characterizes the best-case costs of data structure operations in terms of the component RDMA operations.  The component costs include remote get, remote put, compare-and-swap, and fetch-and-op operations.  We do not distinguish different fetch-and-op operations in this performance model, since the operations involved are typically simple binary functions such as fetch-and-add or fetch-and-XOR, which have very low cost compared to the inherent network latency.  A summary of these operations and the associated notation are shown in Table~\ref{table:component_costs}.

\begin{table}
  \centering
  \begin{tabular}{lll}
    \toprule
    Name & Notation & Latency (us)\\
    \midrule
    put & $W$ & 3.0\\
    get & $R$ & 3.7\\
    compare-and-swap & $A_{\text{CAS}}$ & 3.8\\
    fetch-and-op & $A_{\text{FAO}}$ & 3.9\\
    \bottomrule
  \end{tabular}
  \vspace{1em}
  \caption{Latency of various RDMA operations, measured on Cori with 64 nodes.}
  \label{table:component_costs}
\end{table}

\paragraph*{Alternate Implementations}
As discussed in Section~\ref{sec:pgas_atomicity}, there are different levels of concurrency requirements with which RDMA-based data structure operations can be implemented, depending on the specific needs of an application.  BCL exposes multiple implementations of data structure operations using a mechanism called \textit{concurrency promises}, which allows users to optionally specify the operations that could occur concurrently with the operation being issued.  To illustrate the different levels of concurrency requirements with which a data structure operation could be implemented, consider the case of a hash table insertion with arbitrarily large keys and values.  Inserting an element into such a hash table will, in the general case, require at least two atomic memory operations and a write.  The first atomic memory operation requests a lock on the bucket into which the element will be inserted, the write actually writes the value into the distributed hash table, and a final unlock operation signals that the bucket is ready to be read after the write hash completed.  In this hash table implementation, without the final atomic memory operation, concurrent find operations might read halfway written data, resulting in an incorrect program execution.  However, in the guaranteed absence of concurrent find operations within a barrier region, we can elide the final atomic memory operation, since the following barrier will ensure that the write completes before any find operations may be issued.

Similar levels of concurrency requirements exist for both hash table insert and find operations, as well as operations on queues.  Tables~\ref{table:conpromhash} and \ref{table:conpromqueue} show some of the data structure implementations available in BCL's hash table and queue implementations, along with the associated best case costs.  In the notation used in this paper, $C_\text{W}$ indicates that an operation is allowable with concurrent writes (pushes or inserts), while $C_\text{R}$ indicates that an operation is allowable with concurrent reads (pops or finds) and $C_\text{RW}$ indicates the operation is allowable with either.

% TODO: make these consistent with the notation in this paper
% (In particular the `A` things, which we've added sub-categories for.)
\begin{table}
  \begin{adjustbox}{width=\linewidth}
    \begin{tabular}{| l | l | l | l | l | l |}
      \hline
        Method & Concurrency Level & Description & Cost\\
        \hline
        \textbf{\texttt{insert}}\\
        \hline
        \multicolumn{1}{|r|}{\footnotesize (a)} & Concurrent Read/Write ($C_{RW}$) & Fully Atomic Insert & $A_\text{CAS} + W + A_\text{FAO}$\\
        % \hdashline
        \multicolumn{1}{|r|}{\footnotesize (b)} & Concurrent Write ($C_\text{W}$) & Phasal Insertions & $A_\text{CAS} + W$\\
        \hline
        \textbf{\texttt{find}}\\
        \hline
        \multicolumn{1}{|r|}{\footnotesize (c)} & Concurrent Read/Write ($C_{RW}$) & Fully Atomic Find & $A_\text{FAO} + R + A_\text{FAO}$\\
        % \hdashline
        \multicolumn{1}{|r|}{\footnotesize (d)} & Concurrent Read ($C_\text{R}$) & Phasal Finds & $R$\\
        \hline
      \end{tabular}
      \end{adjustbox}

  \vspace{1em}
  \caption{RDMA-based hash table method implementations considered in this paper.}
  \label{table:conpromhash}
  \vspace{-2em}
\end{table}

\begin{table}
  \begin{adjustbox}{width=\linewidth}
    \begin{tabular}{| l | l | l | l |}
      \hline
        Method & Concurrency Level & Description & Cost\\
        \hline
        \textbf{\texttt{push}}\\
        \hline
        \multicolumn{1}{|r|}{\footnotesize (a)} &  Concurrent Read/Write ($C_{RW}$) & Fully Atomic & $A_\text{FAO} + W + A_\text{CAS-P}$\\
        % \hdashline
        \multicolumn{1}{|r|}{\footnotesize (b)} &  Concurrent Write ($C_\text{R}$) & Only Pushes & $A_\text{FAO} + W$\\
        % \hdashline
        \multicolumn{1}{|r|}{\footnotesize (c)} & Concurrent Local ($C_{\ell}$) & Local Push & $\ell$\\
        \hline
        \textbf{\texttt{pop}}\\
        \hline
        \multicolumn{1}{|r|}{\footnotesize (d)} & Concurrent Read/Write ($C_{RW}$) & Fully Atomic & $A_\text{FAO} + R A_\text{CAS-P}$\\
        % \hdashline
        \multicolumn{1}{|r|}{\footnotesize (e)} & Concurrent Read ($C_\text{R}$) & Only Pops & $A_\text{FAO} + R$\\
        % \hdashline
        \multicolumn{1}{|r|}{\footnotesize (f)} & Concurrent Local ($C_\ell$) & Local Pop & $\ell$\\
        \hline
      \end{tabular}
      \end{adjustbox}

  \vspace{1em}
  \caption{Implementations for circular
           queue methods.}
  \label{table:conpromqueue}
  \vspace{-2em}
\end{table}

% TODO: talk about fine-grained vs. course-grained, buffering.

% TODO: perhaps run this language by Paul/Dan to see if we're
%       using the GASNet verbage correctly. (GASNet vs. GASNet-EX)
% TODO: look up most recent GASNet paper, reference, check their stats.
\subsection{GASNet Active Messages}
GASNet is a communication library that offers remote procedure call functionality in the form of active messages. Active messages are a restricted form of RPC, in that (1) active message handlers cannot require network communication,
% (although they may request it and be denied) Dan suggested we remove this parenthetical, since this is not allowed in the current GASNet-EX impl.
and (2) active message handlers cannot send additional active messages except for request handlers, which may send a single reply to the host.  Since neither of these are necessary for the class data structure operations we consider in this paper and GASNet is known for having a
% TODO: is there a less-biased sounding way of saying this?
high-quality, fast implementation of active messages, we use GASNet to implement a set of RPC-based distributed data structure implementations to compare against BCL's RDMA-based data structures.

In our data structure implementations, we use GASNet-EX 2019.6.0, the most recent version of GASNet-EX API at the time of submission, and our discussion of the active messages API are as present in this version of GASNet-EX.  GASNet-EX active messages consist of a fixed-size header, which includes an index referencing the desired handler function, and up to 64 bytes of arguments, along with an optional variable length payload.  Active message handlers must be registered with the GASNet runtime before they can be used.  When an origin process wishes to invoke an active message on a remote target process, it issues an active message request.  A target process, inside the context of an active message handler, can optionally issue an active message reply to the origin process, which will result in the corresponding reply handler running on the origin process.  To wait for the completion of an individual active message, an origin process must wait until a reply handler issued by the target process has finished running locally, writing some reply data or otherwise indicating that it has completed.

% TODO: either implement this, or else justify why we're not using this.
%       (Essentially, this is just a performance optimization for
%        one-sided only ops that don't need individual completion.)
%       Dan suggested that the fastest confirmation mechanism is just sending a
%       reply, as it removes the need for collective synchronization.
%A collective flush, guaranteeing the completion of all active messages, is implemented by having each process keep track of the number of active messages it has requested and have received a reply.  To perform a flush, each process individually waits in a poll loop until those two numbers are equal, then enters a global barrier.

In order to service active messages, each process must enter the GASNet AM poll function, which can be called by the main process or from a progress thread that constantly checks the queue in order to provide attentiveness.

\section{Experimental Design}
In this section, we examine the performance of RDMA- and RPC-based designs for remote operations on distributed data structures.  Our expectation is that there are some data structures, applications, and workloads for which RPC, with its greater expressiveness, will achieve higher performance, and some situations where RDMA, with its lower round trip latency and hardware-accelerated execution in the network interface card, will achieve higher performance.  First, we perform microbenchmarks to measure the \textit{component costs} for RDMA- and RPC-based distributed data structure implementations.  That is, the cost of a remote put, remote get, compare-and-swap, and fetch-and-op operations, as well as the round trip cost of sending a GASNet active message request and receiving a reply.

% XXX: thought---if we implement the collective-based flush,
%      we'll have two component benchmarks for AM instead of
%      one, which will look better.

%\section{Motivation}
%The main objective in this paper is to investigate the trade-off between an RDMA and AM on the application abstraction level. Many work focused exclusively on using RDMA reads in their distributed data structure to traverse remote data structures, similar to what would have been done had the structure been in local memory. This is a natural way since it is very alike its shared memory version and one-sided RDMA operations are efficient and desired in terms of its latency.  However, a challenge for RDMA lies in the lack of richness of RDMA operations. An RDMA operation can only read or write a remote memory location. To illustrate this, consider an example of implementing a distributed hash-table’s insert function. Should insert be implemented by having the requester repeatedly attempt (in case of collisions) to directly inserting a (key, value) pair into the shared hash table on a remote node using RDMA operations, or should insert be implemented by using AM to send the (key, value) pair, along with the insert function id, to the remote node, and letting the remote CPU perform the repeated probing attempts locally? The former approach uses direct RDMA operations which, as noted earlier, have lower latency and higher throughput. The latter approach, however, avoids the need to make multiple round trips upon probing failures.

\subsection{Component Benchmarks}
For the component benchmarks, we measure the cost of small-size remote put and remote get operations, along with the cost of performing the atomic compare-and-swap fetch-and-op operations.  In each of these microbenchmarks, we begin with a globally visible array located in the shared segment of each process in the program.  To perform the benchmark, each process will continuously perform a remote memory operation to a random location in a random process' globally visible array.  After the benchmark completes, we divide the total amount of time taken by the number of operations completed per process to arrive at the latency of the individual operations.

We also measure the cost of sending an active message.  In this case, each process continually sends an active message to another random process and then waits for a reply to complete before proceeding.

The purpose of collecting these benchmarks is not only to evaluate the relative costs of the operations that make up RDMA-based distributed data structure methods, but also to evaluate BCL's analytical performance model.  By plugging in the component method costs into the formulas for different data structure operations, we can evaluate to what extent observed performance deviates from theoretical best case performance.  Performance could deviate for a number of reasons, including higher than optimal latency due to specific application workloads stressing the network hardware and high contention leading to many more round trips than would be necessary in the best case.  This analysis will allow us to evaluate what makes up the cost of RDMA-based remote data structure operations, which can both allow data structure developers to better design data structure operations, prioritizing use of cheaper component operations, and allow hardware designers to identify which RDMA operations to optimize in order to increase RDMA-based data structure performance.

\subsection{Data Structure Benchmarks}
After collecting microbenchmark results, we ran a series of experiments where we benchmarked different distributed data structure operations in BCL and compared them with equivalent active message implementations.  Full descriptions of these distributed data structure implementations can be found in the original BCL paper~\cite{Brock:2019:BCD:3337821.3337912}.

% TODO: perhaps list the precise notation for each one of these in the text,
%       point to the table.
\subsubsection{Hash Table}
Distributed hash tables are an important data structure for many applications, including various data analysis problems such as genomics.  In BCL, hash tables are implemented as a distributed array of buckets, where each bucket contains room for a key, a value, and a flag that will be used for synchronization.
\paragraph*{$C_\text{RW}$ Insert}
As discussed earlier, to achieve fully concurrent safety, a hash table insertion requires, in the best case, two atomic memory operations and a remote put operation.  In the BCL implementation, this is a compare-and-swap operation to request the hash table bucket, a remote put, and an atomic fetch-and-AND operation to mark the bucket as ready to read.
\paragraph*{$C_\text{W}$ Insert}
An insert operation without find concurrently occurring can be preformed using using only one atomic memory operation, setting the bucket's flag to ready, then performing a remote put.  The collective barrier that must separate the inserts from any find operations will guarantee that the remote put has completed before any find operations can read the value.
\paragraph*{$C_\text{RW}$ Find}
To perform a fully concurrently safe find operation, the reading process must first obtain a read lock on a bucket before reading what is inside it.  This is to prevent other processes from modifying the bucket while the process is reading it.  In BCL's implementation, this is done with an atomic fetch-and-OR operation to set one of a number of read bits in the flag.  After the lock is obtained, the process will read the bucket with a remote get operation, then unset the read bit with an atomic fetch-and-AND.
\paragraph*{$C_\text{R}$ Find}
A find without inserts concurrently occurring can be performed with a single remote get operation, since it is able to retrieve both the flag and the key and value in a single remote memory operation.

\subsubsection{Queue}
Queues are widely used data structures in many applications such as data redistribution and asynchronous all-to-all operations, producer-consumer problems, frontier based graph algorithms, and others.
% TODO: add references.
The most crucial operations for queues are pushing and popping.  Since these operations are symmetric, we only consider push operations here for reasons of brevity.
Queues in BCL are implemented as \textit{hosted} data structures, meaning they live on a single host process, but are visible to all processes.  Applications may either require a single queue to be manipulated by all or a subset of processes, or, commonly, a queue on each process.  Queues are implemented as a ring buffer, with sets of head and tail pointers marking the beginning and end of data stored within the queue.
\paragraph*{$C_\text{RW}$ Push}
Similar to the hash table insertion example, with this queue implementation, a fully concurrently safe push requires two atomic memory operations and a remote put operation.  The first operation is a fetch-and-add operation, which requests space in the queue by advancing the tail pointer, followed by a remote put to the reserved space.  Finally, a remote compare-and-swap operation is necessary to advance the ready tail pointer to indicate that the written segment of the queue is ready to be read.  A fetch-and-add operation is not correct here, since it could advance the ready tail pointer past previous insertions which are not yet finished writing.
\paragraph*{$C_\text{W}$ Push}
A push without pops concurrently occurring can be completed with a single atomic fetch-and-add, to reserve space, followed by a remote put to write to the reserved spot in the queue.  This is because a barrier will separate all the push operations from any pop operations, thus guaranteeing that the written data is ready to be read.

\subsection{RPC Implementations}
Our RPC implementations, based on GASNet-EX active messages, consist of a handler function, which performs the relevant operation on a local data structure then sends a reply.  If the operation has no return value, the reply will simply increment a counter on the origin side, which can be used to ensure fine-grained completion of data structure operations.  If the operation has a return value, it will also write the return value to a memory location passed in by the request AM.

\section{Results}

First we measured our set of \textit{component costs}, which include individual operations that make up remote data structure operations.  Measured operations include 32-bit remote get, remote put, atomic fetch-and-add, atomic compare-and-swap, and a round-trip active message with a payload of 64-bits.  Each process picks a random process for each operation, and, for the case of the RDMA operations, a random memory location.  The active message experiment measures the cost of a round trip with a 64-bit payload, with the inner operation being an insertion into a remote hash table.
\paragraph*{Experimental Setup}
We measured the component costs on Cori Phase I, which is a Cray XC40 supercomputer with a Cray Aries interconnect and 32 cores per node.  All benchmarks were run with one process per core.  Experiments were run with 100,000 local elements per process, unless noted otherwise.  In each experiment, the target operation is executed a million times inside a loop, then the total time spent inside the loop is divided by the number of issued operations to calculate the operation's latency.  In order to avoid systematic errors due to variations in network performance, which may be caused by varying job placement within a cluster as well as contention with other jobs, each dot on each graph was submitted as a separate job submission, averaged over at least four jobs.

\begin{figure}
  \includegraphics[width=\linewidth]{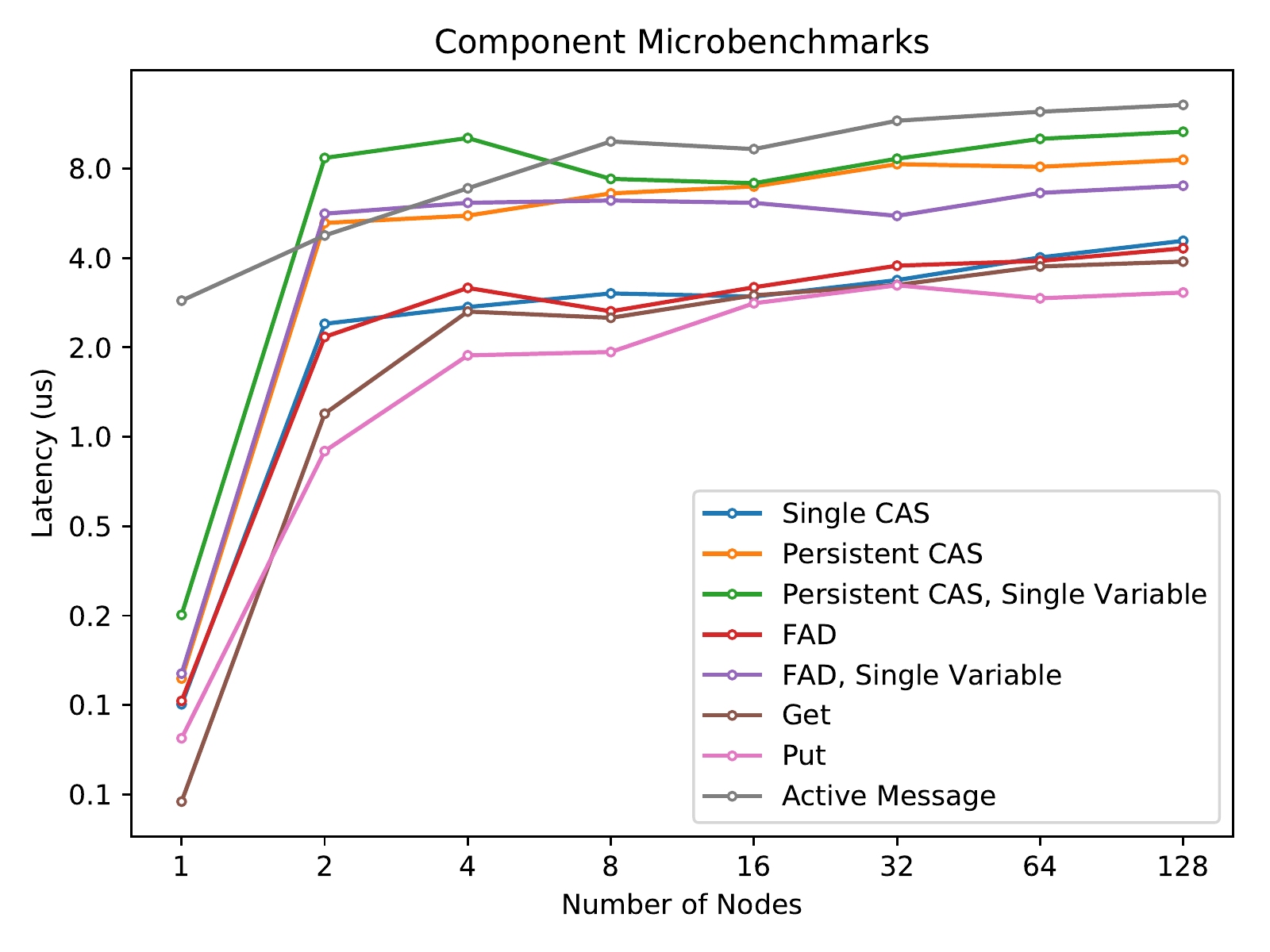}
  \caption{The component latencies for RDMA operations and AMs on Cori.}
  \label{fig:component_latency}
\end{figure}

\paragraph*{Component Benchmarks}
Our component benchmark results are shown in Figure~\ref{fig:component_latency}.  This figure includes the benchmarks discussed above, along with two extra versions for the two measured atomic operations.
% TODO: remove Persistent CAS, Single Variable from Graph
\paragraph*{Compare-And-Swap}
We show two benchmarks for compare-and-swap, ``Single CAS,'' which measures the cost of a single compare-and-swap operation and ``Persistent CAS,'' which measures the cost of a compare-and-swap which continually polls until it succeeds in changing the value from the previous value to a new value.  The Single CAS experiment measures the cost of the CAS AMO operation, while the Persistent CAS experiment provides some measure of the cost of a persistent CAS that repeatedly polls until it succeeds in modifying the target value.
\paragraph*{Fetch-And-Add}
We include two versions of the fetch-and-add operation, ``FAD,'' which measures the cost of a fetch-and-add operation issued to a random value on another process, and ``FAD, Single Variable,'' which measures the cost of a fetch-and-add operation issued when there is only a single target variable per process.

In general, there is a large jump in latency for RDMA operations when moving from a single node to two nodes, which is to be expected when switching from shared memory to distributed memory.  From there, each operation increases in cost gradually, which can be explained by (1)~the decreasing percentage of operations that will be operating on local, fast memory and (2)~an increase in the distance that messages must travel across the network as the allocation size increases.

We find that the Put operation has the lowest cost, followed by a cluster of operations of similar cost, including Get, FAD, and Single CAS.  As one might expect, the ``Persistent CAS'' operations are much more expensive, since they may require multiple round trips to succeed in swapping the target value.  More surprisingly, we also find that the ``Single FAD'' benchmark, which operates on a single target value per process, has a much higher cost than the ``FAD'' benchmark, which operates on a range of values per process.  This indicates that, for this operation, the \textit{target memory locations}, not just the number of incoming operations on the target NIC, can impact the amount of time that a fetch-and-add operation will take (at least on the Cray Aries NIC).  While this might be expected for a shared memory environment, where a directory or snooping protocol must be used to ensure cache coherence, NIC-accelerated atomic memory operations are not atomic with respect to CPU atomics, and the authors expected the speed of NIC-accelerated fetch-and-add atomics to be unaffected by the target address.  Indeed, experiments on the Summit supercomputer, which has an Infiniband FDR interconnect, did not reveal any difference between the two benchmarks.

% Commenting this out for the moment --- we can chat about whether to
% expand Table I to include these stats or not, but we probably don't need
% 2 copies of it.
%\begin{table}
%  \centering
%  \begin{tabular}{lll}
%    \toprule
%    Operation & Latency(us) & Cost(AM) \\
%    \midrule
%    active message   & 12.4 & 1.0  \\
%    put              & 2.9 & 0.24 \\
%    get              & 3.7 & 0.30 \\
%    compare-and-swap & 4.0 & 0.32 \\
%    persistent CAS   & 8.1 & 0.65 \\
%    persistent CAS(single) & 10.1 & 0.81 \\
%    fetch-and-op     & 3.9 & 0.31 \\
%    fetch-and-op(single)   & 6.6 & 0.53 \\
%    \bottomrule
%  \end{tabular}
%  \vspace{1em}
%  \caption{Analytical performance model for RDMA operations on Cori, 64 nodes.}
%  \label{table:analytical_performance_model}
%\end{table}

% TODO: figure out how we're going to reorganize this graph.
% XXX: Leave this graph out for now
%\begin{figure}
%  \includegraphics[width=\linewidth]{contention.pdf}
%  \caption{The component latencies for RDMA operations with different level of %contention on Cori.}
%  \label{fig:contention_latency}
%\end{figure}

\begin{figure}
  \centering
  \includegraphics[width=\linewidth]{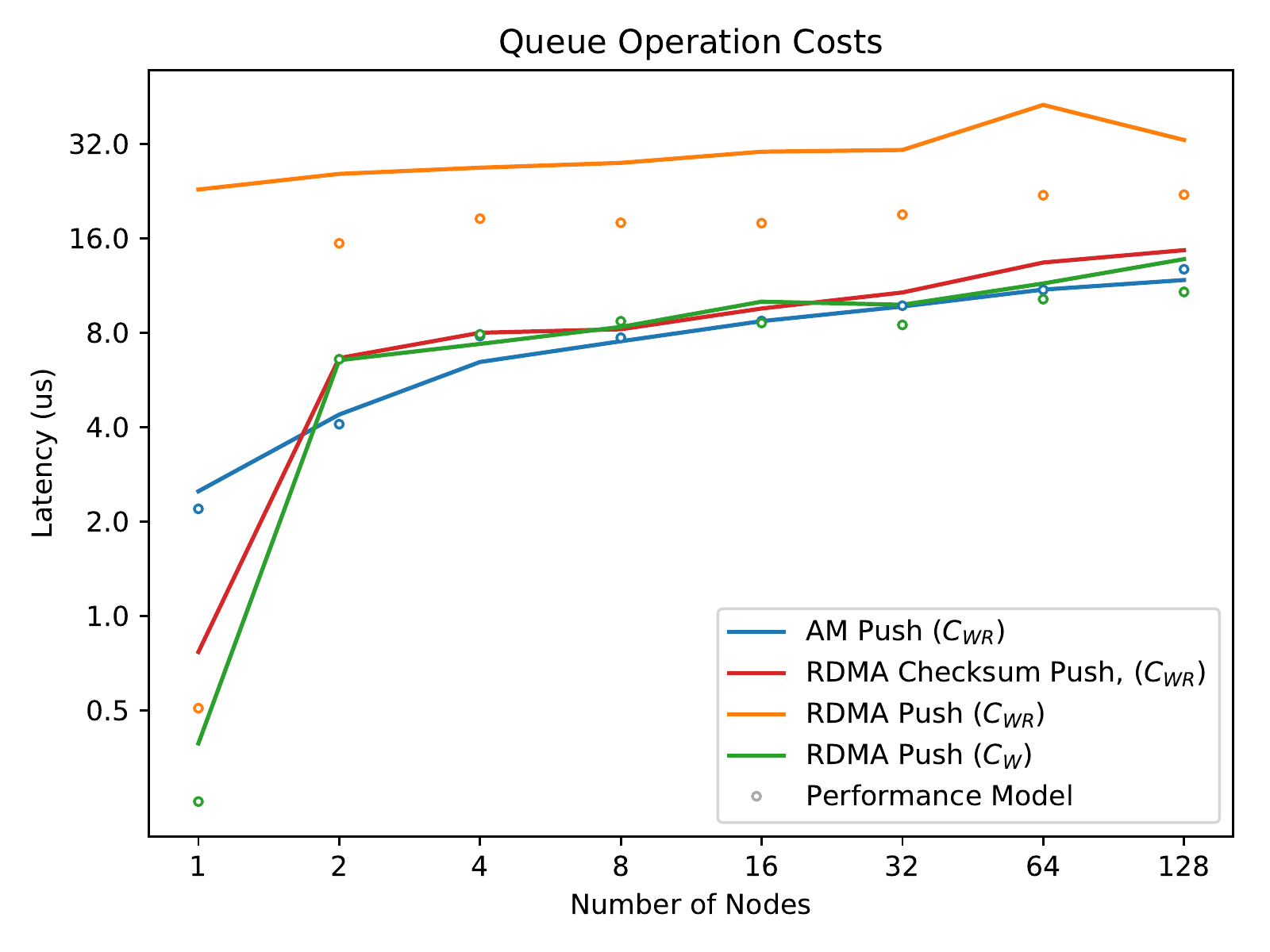}
  \caption{Latencies for RDMA- and RPC-based queue push operations.}
  \label{fig:queue_latency}
\end{figure}

\paragraph*{Queue Benchmarks}
Next, we measured the cost of a set of \textit{queue data structure} microbenchmarks, and compared our empirical data structure benchmark results with our performance model's prediction using the component microbenchmark results.  These results are shown in Figure~\ref{fig:queue_latency}.  Pushing and popping are the primary queue operations, and since they are symmetric, we choose to only show queue push results here for reasons of space.  We compare four different queue results: (1) an ``AM Push'' benchmark, which uses an active message to insert into a local queue on each process, (2) a ``RDMA Push $C_\text{W}$'' benchmark, which is a phasal queue that allows either concurrent pushing or popping, but not simultaneously, (3) an ``RDMA Push $C_\text{RW}$'' benchmark, which uses an additional atomic operation to signal the completion of the write of data onto the queue, and (4) a ``RDMA Checksum Push $C_\text{RW}$'' benchmark, which is a separate design of queue that in addition to writing the data values into the queue also writes a checksum value that can be used to verify whether the write has finished.
% TODO: clarify the whole thing about how queue synchronization works.

After accounting for the increased cost of the ``Single FAD'' experimental result and plugging that in as the parameters for our analytical performance model in Table~\ref{table:component_costs}, we found that the performance model generally predicted the behavior of the queue data structure benchmarks.  The one exception is the RDMA Push ($C_\text{RW}$) benchmark, which is consistently more expensive than the performance model would predict.  This appears to be due to the fact that the second atomic memory operation in the $C_\text{RW}$ push operation requires more round trip attempts than the Persistent CAS microbenchmark would suggest.  While our Persistent CAS microbenchmark attempts to change the target value from the previously seen value to the desired value, the persistent CAS involved in the queue benchmark attempts to increase a ``tail ready'' pointer that marks the frontier of values written into the queue.  It may only proceed after any other insertions have finished writing, which leads to some inherent serialization that is not represented in the performance model.

\begin{figure}
  \centering
  \includegraphics[width=\linewidth]{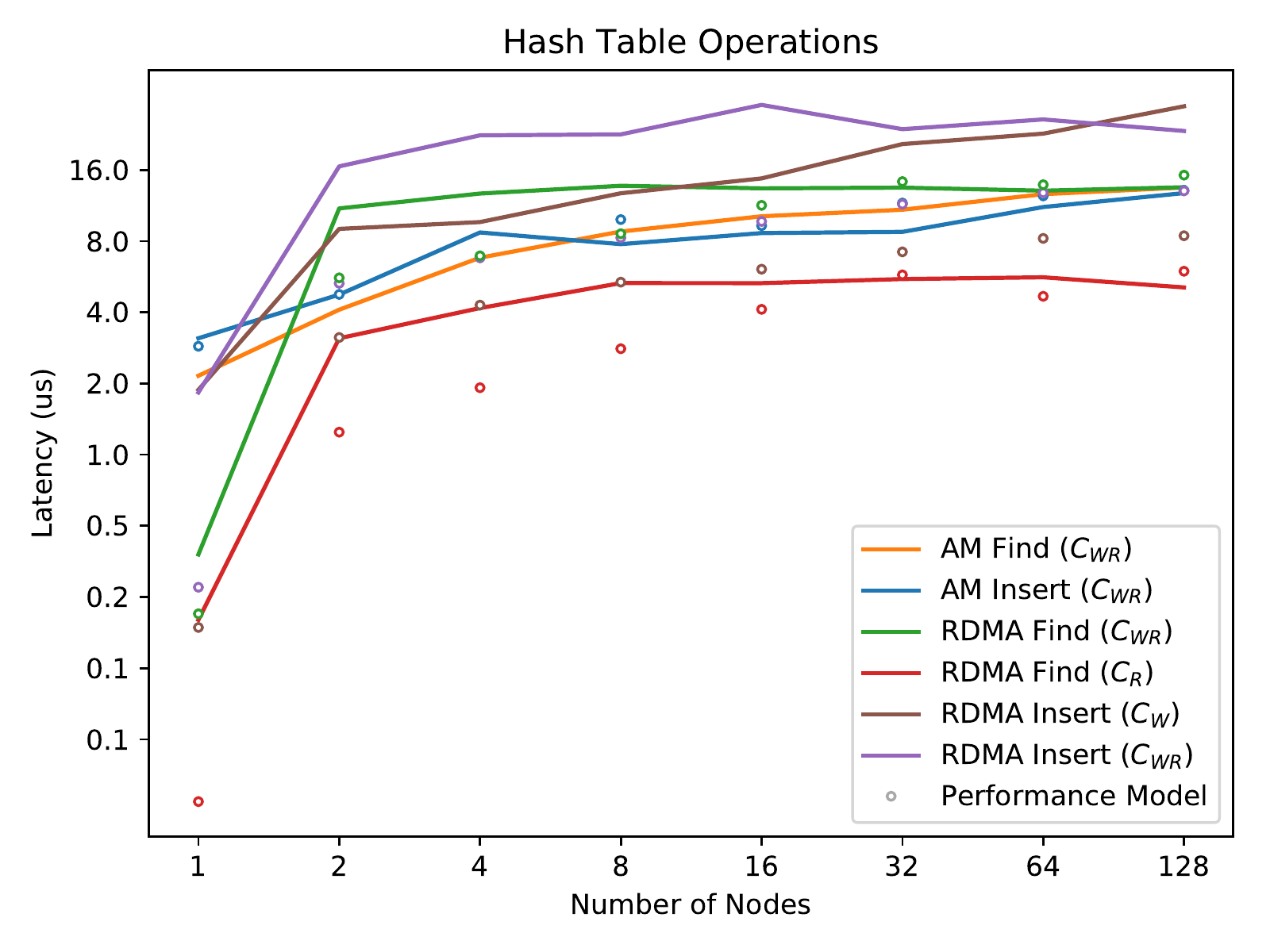}
  \caption{Latencies for RDMA- and RPC-based hash table operations.}
  \label{fig:hash_latency}
\end{figure}

\paragraph*{Hash Table Benchmarks}
We also measured the cost of several hash table data structure operations, which are displayed in Figure~\ref{fig:hash_latency}.  The less expensive RDMA Find ($C_\text{R}$) operation is the cheapest operation, followed by the active message implementations of AM Find ($C_\text{RW}$) and AM Insert ($C_\text{RW}$), with find possibly having a slightly higher cost, due to the fact that the return trip message is slightly larger, containing a return value.  The more expensive RDMA Find ($C_\text{RW}$) is initially slightly more expensive than the active message implementations, but appears to scale better, ending up at a similar cost at 128 nodes.  The insert implementations, both $C_\text{RW}$ and $C_\text{W}$, are more expensive.  Both RDMA operations seem to roughly match their associated performance models, with some increase in the real benchmark runtime perhaps due to hash table collisions, which are not included in the performance model.  Surprisingly, both hash table insertion methods vary significantly from their associated cost models' prediction.  However, except for RDMA Insert $C_{W}$, the predicted order of implementations in terms of performance is correct.
% TODO: expand this section, try to explain discrepancies.

\begin{figure}
    \centering
    \includegraphics[width=\linewidth]{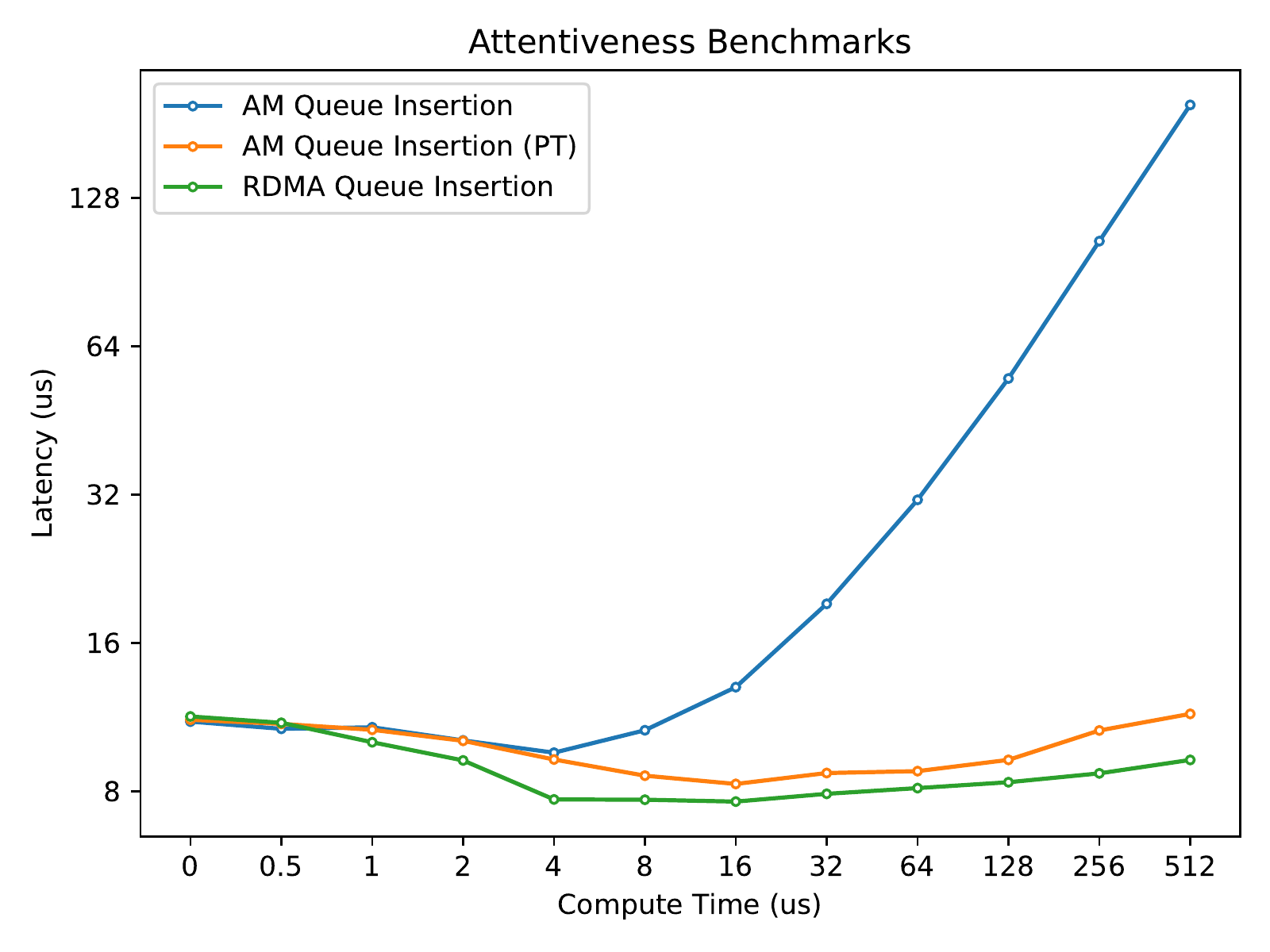}
    \caption{Measuring the cost of a queue insertion as remote processes become less attentive due to intermixed computation.}
    \label{fig:attentive}
\end{figure}

\paragraph*{Attentiveness Benchmarks}
Each of the above active message benchmarks could be considered a close to optimal case in terms of \textit{attentiveness}, by which we mean the availability of remote processes to service active message requests.  This is because, without an independent progress thread to ensure attentiveness, which is the model assumed in the above benchmarks, a process must enter a progress function in order to ensure that inbound requests are serviced.  In each of the above benchmarks, each process issues a single active message request, then polls on a progress function, servicing incoming active messages, until a reply is received for the active message request.  In a more realistic scenario, remote data structure operations will be interspersed with computation, which may impact the attentiveness of remote processes, resulting in longer latencies for active messages.  Figure~\ref{fig:attentive} shows the impact of adding interspersed computation on the latency of a queue insertion.  We arrived at this plot by inserting a small function to perform a given number of microseconds of computation inside a loop of queue insertions.  To calculate only the time spent performing queue insertions, we subtract the compute time from the total time taken in the calculation.  As shown in the plot, the active message version, while initially faster than the RDMA-based implementation, quickly becomes more expensive as the interspersed computation time exceeds 2 microseconds.  From there, the increase in queue insert time grows roughly linearly with the compute time, as the average time an active message must wait to be serviced is half the compute time.  We observe that the latency for the RDMA queue insertion actually goes down, which we attribute to lower latency across a quieter network as queue insertions are spaced out among a greater quantity of computation.  It is important to note that RPC-based implementations can attain better attentiveness by explicitly using a progress thread, which will continually poll for new RPC operations to execute.  In the figure, ``AM Queue Insertion (PT)'' demonstrates this.  When using a progress thread, active messages are not subject to the same pathological behavior due to lack of attentiveness, but do receive a performance hit, likely due to contention between the progress and main compute thread.

%\begin{figure}
%  \includegraphics[width=\linewidth]{latency_am_rdma_summit.pdf}
%  \caption{The component latencies for RDMA operations and AMs on Summit.}
%  \label{fig:contention_latency_s}
%\end{figure}

%\begin{figure}
%  \includegraphics[width=\linewidth]{contention_summit.pdf}
%  \caption{The component latencies for RDMA operations with different level of contention on Summit.}
%  \label{fig:contention_latency_s}
%\end{figure}
%\section{Discussion}

\section{Related Work}
UPC++ is a high-level asynchronous PGAS programming interface for C++~\cite{bachan2019upc++}.  UPC++ has a heavy emphasis on asynchrony and allows users to build arbitrary computation graphs by combining futures together with callbacks.  UPC++ also encourages users to build data structures and applications on top of RPC.

AM++ is C++ library built on top of MPI that provides a high-level active message API similar to that discussed in this paper~\cite{willcock-amplusplus}.  Active Pebbles extends AM++ by adding support for message aggregation and more sophistocated termination detection mechanisms~\cite{active-pebbles}.  You've Got Mail (YGM) is an MPI-based system that provides an active message-like API with message- and node-level aggregation~\cite{priest2019you}.

% TODO: add Conveyors?
%Also \cite{bale2019}.

STAPL is a parallel programming framework that provides distributed data structures loosely based on a partitioned global address space (PGAS) model, but is not strictly designed to support RDMA~\cite{Tanase:2011:SPC:1941553.1941586}.  PETSc, Chombo, and AMReX provide data structures for sparse and dense matrices and structured and unstructured grids, but do not focus on the types of irregular, generic data structures discussed here~\cite{abhyankar2018petsc,colella2009chombo,AMReX_JOSS}.

Aguilera, et al. have proposed various hardware extensions to RDMA specifically to allow for the efficient execution of operations on remote data structures in NIC hardware~\cite{Aguilera:2019:DFM:3317550.3321433}.  These include an indirect access primitive that can be used to access a value at an offset from a pointer on a remote node, allowing for a dynamically resizing remote vector; various scatter and gather primitives, and a form of notifications.  We believe that performance models like the one presented here are a good fit for evaluating potential new RDMA instructions, and that similar microbenchmark analysis can help hardware and software developers to design and evaluate new hardware and data structures.

\section{Conclusions}
In this paper, we compared implementing distributed data structures using RDMA and RPC.  We developed an analytical performance model which predicts the performance of the distributed data structures based on their components, then compared this to real-world performance. In most of the cases, our model's predictions matched the real-world results.  We observed the impact of system-specific hardware behavior, namely the increased cost of a fetch-and-add performed on a single memory location on Cray Aries; and also observed the impact of increased contention due to multiple round trips, as in the case of a concurrent read/write queue insertion.  We also examined the impact of attentiveness on RPC performance, observing that RDMA may have advantages when it comes to communication interspersed with computation.

%\section*{Acknowledgments}
%All the grants go here. [TODO]

\bibliographystyle{./bibliography/IEEEtran}
\bibliography{bcl-refs}

\clearpage
\appendices
\section{Artifact Description: RDMA vs.~RPC for Implementing Distributed Data Structures}

\subsection{Abstract}
This artifact describes all experiments presented in the submission titled ``RDMA vs.~RPC for Implementing Distributed Data Structures.''

\subsection{Description}
\subsubsection{Check-list (artifact meta information)}
\begin{itemize}
  \item \textbf{Program} C++, BCL, and GASNet-EX.
  \item \textbf{Compilation} Compilation with GCC 8.2.0 and GASNet-EX 2019.6.0
  \item \textbf{Datasets} No datasets.
  \item \textbf{Runtime Environment} SUSE Linux Enterprise Server 12 on NERSC Cori.
  \item \textbf{Hardware}
  \begin{itemize}
    \item NERSC's Cori Phase I supercomputer, a Cray XC40.  Each node is equipped
    with two 16-core Intel Xeon E5-2698 v3 CPUs, which use the Intel Haswell
    microarchitecture.  All nodes have 128 GB of RAM and are connected via a
    Cray Aries interconnect.
    \item NERSC's Cori Phase II supercomputer, a Cray XC40.
    Each node is a self-hosted 68-core Intel Xeon Phi 7250 accelerator, which
    uses the Intel Knights Landing microarchitecture.  All nodes are equipped
    with 96 GB of DRAM and 16 GB of MCDRAM and are connected via a Cray Aries
    interconnect.
  \end{itemize}
  \item \textbf{Output} Total execution time and operation latency (in seconds and microseconds).
  \item \textbf{Experiment Workflow} Download software, compile from source, run
  the applications, examine the outputs.
  \item \textbf{Publicly available?} Yes.
\end{itemize}

\subsubsection{How software can be obtained}
\begin{itemize}
  \item BCL can be downloaded from \url{https://github.com/berkeley-container-library/bcl}.
  All benchmarks can be found in the \texttt{examples/benchmarks/am-comp} directory on branch \texttt{am-comp}.
  \item GASNet-EX can be downloaded from \url{https://gasnet.lbl.gov/}.
\end{itemize}

\subsubsection{Hardware dependencies}
The experiments can be performed on any cluster than supports RDMA, although the results may have discrepancies due to differences in hardware, which is expected.  Our precise results should be reproducible on Cray Aries systems similar to Cori Phase I.

\subsubsection{Software dependencies}
To compile and run our benchmarks, a C++-17-compliant compiler along with a copy of GASNet-EX is required.  Reproducing our precise results will also require that GASNet-EX be compiled with the Cray uGNI conduit for the Cray Aries interconnect.

\subsubsection{Datasets}
No datasets are required to run our benchmarks.

\subsection{Installation}
\begin{itemize}
  \item BCL can be installed by cloning the BCL Git repository at
  \url{https://github.com/berkeley-container-library/bcl}, checking out branch \texttt{am-comp}, and adding the new directory to the \texttt{CPLUS\_INCLUDE\_PATH} environment variable.
  \item Instructions for installing GASNet-EX are available at \url{https://gasnet.lbl.gov/}.  Our BCL Makefiles require that the \texttt{gasnet\_prefix} environment variable be set to the directory of the GASNet-EX installation.
\end{itemize}

\subsection{Experiment Workflow}
Each benchmark is available in the \texttt{examples/benchmarks/am-comp} folder of the \texttt{am-comp} branch of the BCL Git repository.  Each set of benchmarks can be compiled with the Makefile included in the corresponding directory.  It may be necessary to modify the Makefile if it is necessary to use another conduit or if the user moves the benchmark files out of the main BCL directory.

To reproduce our results, each benchmark should be executed with one process per core, or 32 processes per node on Cori Phase I.  This can be done with Slurm commands of the form \texttt{srun -N NNODES -n \$((NNODES*32)) ./benchmark}.  To modify the \textit{local size}, which is the size of each process' local portion of the global array in the components benchmarks, users may use the flag \texttt{-s}.  To change the number of operations performed in each benchmark---for example because a particular operation has a different cost on a particular architecture and thus the total benchmark runtime is unreasonably fast or slow---users may set the \texttt{-n} flag to control the number of operations performed.

\end{document}